# Design and Implementation of a Thomson Parabola for Fluence Dependent Energy-Loss Measurements at the Neutralized Drift Compression eXperiment

F. Treffert,[1,2,a] Q. Ji,[1] P.A. Seidl,[1] A. Persaud,[1] B. Ludewigt,[1] J.J. Barnard,[3] A. Friedman,[3] D.P. Grote,[3] E.P. Gilson,[4] I.D. Kaganovich,[4] A. Stepanov[4], M. Roth[2] and T. Schenkel[1]

[1]*Lawrence Berkeley National Laboratory, 1 Cyclotron Road, Berkeley, California, 94720, USA*

[2]*Department of Nuclear Physics, Technical University Darmstadt, Schloßgartenstraße 9, 64289, Darmstadt, Germany*

[3]*Lawrence Livermore National Laboratory, 7000 East Avenue, Livermore, California, 94550, USA*

[4]*Princeton Plasma Physics Laboratory, 100 Stellarator Road, Princeton, New Jersey, 08540, USA*

The interaction of ion beams with matter includes the investigation of the basic principles of ion stopping in heated materials. An unsolved question is the effect of different, especially higher, ion beam fluences on ion stopping in solid targets. This is relevant in applications such as in fusion sciences. To address this question, a Thomson parabola was built for the Neutralized Drift Compression eXperiment (NDCX-II) for ion energy-loss measurements at different ion beam fluences. The linear induction accelerator NDCX-II delivers 2 ns short, intense ion pulses, up to several tens of nC/pulse, or $10^{10}$-$10^{11}$ ions, with a peak kinetic energy of ~1.1 MeV and a minimal spot size of 2 mm FWHM. For this particular accelerator the energy determination with conventional beam diagnostics, for example, time of flight measurements, is imprecise due to the non-trivial longitudinal phase space of the beam. In contrast, a Thomson parabola is well suited to reliably determine the beam energy distribution. The Thomson parabola differentiates charged particles by energy and charge-to-mass ratio, through deflection of charged particles by electric and magnetic fields. During first proof-of-principle experiments, we achieved to reproduce the average initial helium beam energy as predicted by computer simulations with a deviation of only 1.4 %. Successful energy-loss measurements with 1 μm thick Silicon Nitride foils show the suitability of the accelerator for such experiments. The initial ion energy was determined during a primary measurement without a target, while a second measurement, incorporating the target, was used to determine the transmitted energy. The energy-loss was then determined as the difference between the two energies.

## I. INTRODUCTION

Stopping power measurements investigate the effect radiation has on matter and possible damage that occurs when materials are heated through deposition of energy by ions passing through the material. The mechanisms of ion stopping in matter are well understood nowadays except for the effect of different, especially higher ion fluences on ion stopping in matter. Such a fluence dependent effect is detectable by monitoring the kinetic energy of the transmitted ion beam. NDCX-II is a facility that enables such experiments. The linear induction accelerator NDCX-II is designed to deliver short, intense helium ion pulses[1], up to several tens of nC/pulse, or $10^{10}$ - $10^{11}$ ions, with a peak kinetic energy of approximately 1.1 MeV and a repetition rate of one shot every 45 s. The ~12 m long accelerator consists of a helium plasma ion source driven by a filament[2], induction cells for particle acceleration and axial compression, diagnostic cells and a neutralized drift compression section[3]. Solenoids, with a maximum magnetic field of 2.5 T, in each cell steer and focus the beam radially. In 12 induction cells the beam is accelerated and rapidly compressed with voltage waveforms whose slopes and amplitudes can be adjusted individually. Peak voltages range from 15 kV to 200 kV with durations of

a) current address: SLAC National Accelerator Laboratory, 2575 Sand Hill Rd., Menlo Park, CA 94025, USA

0.07 - 1 µs for the different induction cells[1]. After the last acceleration gap, an imparted head-to-tail velocity ramp further compresses the beam to a pulse length of 2 ns FWHM as it coasts to the target position. During the coasting, the beam passes through a preformed plasma in the drift compression section to counteract the significant space charge repulsion from the self-fields of the beam bunch. The final focus solenoid (max. field 8 T) radially focuses the beam on the target to a minimum spot size diameter of 2 mm. Target temperatures of up to ~ 0.1 eV, dependent on target properties, can be obtained through Bragg Peak heating with currently achieved beam properties[4]. The beam behavior in the accelerator is simulated using the particle-in-cell (PIC) code WARP[5]. The predictions of the arrival time of the beam from WARP are in fair agreement with experiments, which leads to the more detailed question of the accuracy of the WARP energy distribution of the beam. A Thomson parabola[6-8] was designed and implemented to accurately measure the beam particle energies for studies of ion solid interactions. The first proof-of-principle energy-loss measurements at NDCX-II using 1 µm thick Silicon Nitride foils were performed using the designed Thomson parabola.

## II. THOMSON PARABOLA DESIGN

The Thomson parabola is positioned 47.2 cm ($d_{T-P}$) downstream from the target, as shown in Figure 1. It consists of a pinhole ($d_2$ = 0.4 mm) attached to a shielding plate, ~12 cm x 10 cm large and 16 mm thick stainless steel, which blocks background light produced by the source and the drift compression section. Five centimeters behind the plate superimposed magnetic (red) and electric (green) fields deflect the beam with geometric lengths $L_B$ = 9.5 cm (magnetic) and $L_E$ = 15.3 cm (electric). A drift section of $D_E$ = 15 cm separates the end of the electric plates from the detector, a polyvinyltoluene scintillator (Saint Gobain BC-408)[9], 10 cm diameter and ~5 mm thick, and an image-intensified gated CCD camera (Princeton Scientific PIMAX 2) with a pixel size of 19 x 19 µm and 512 x 512 pixels. In front of the target an additional pinhole ($d_1$ = 2 mm) is added to the setup. The two pinholes lead to a selection of ions with angles below 2.5 mrad adding an error of approximately 2 % to the energy determination.

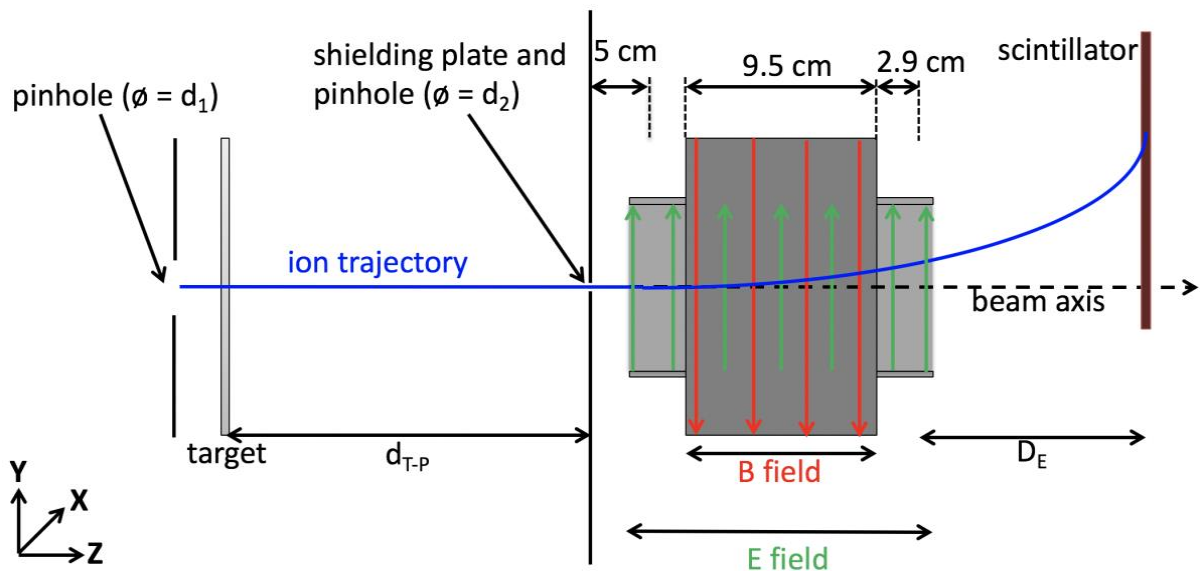

FIG. 1. The Thomson parabola is located 47.2 cm ($d_{T-P}$) after the target. It consists of a shielding plate with a 0.4 mm ($d_2$) diameter pinhole, located 5 cm in front of the magnet assembly. The assembly consists of a $L_B$ = 9.5 cm long magnet and $L_E$ = 15.3 cm long electric plates. The drift between the end of the electric plates and scintillator measures 15 cm ($D_E$). Another pinhole ($d_1$ = 2 mm) is located directly in front of the target.

Figure 2 shows the magnet and electric plate assembly (B - E assembly) as seen along the beam axis looking downstream from the target. The assembly is positioned on a tray which is fixed to two 1D stages, enabling precise positioning in the transverse (x-y) direction. The ions deflected by the assembly hit the plastic scintillator (green) in front of a viewing window. The CCD camera behind the window detects the emitted scintillator light and is gated within a 100 ns window around the beam pulse for background rejection but allowing the capture of the tail of the beam. A slider stage fixed to the flange at the back of the target chamber makes the scintillator movable. Moving the scintillator and B - E assembly to the "out" position enables the investigations of ion beam properties at the target position for beam tuning purposes. Stacked neodymium (NdFeB) magnets create a permanent magnetic field in negative y-direction. The electric field, oriented in positive y-direction, is created in between two parallel aluminum plates by applying opposite polarity voltages to the plates. By varying the applied direct current (DC) voltages it is possible to vary the electric field occurring inbetween the plates. The antiparallel fields cause a positive y-deflection (electric field), and a positive x-deflection (magnetic field) according to the Lorentz force. The resulting magnetic ($x_{mag}$) and electric ($y_{el}$) deflection equations

$$y_{el} = \frac{q \, e \, E \, L_E \left(D_E + \frac{L_E}{2}\right)}{2 \, E_{kin}}, \quad (1)$$

$$x_{mag} = \sqrt{\frac{q^2 \, e^2 \, B^2 \, L_B^2 \left(D_B + \frac{L_B}{2}\right)^2}{2 \, m \, E_{kin}}}, \quad (2)$$

are dependent on the field strengths $B$ and $E$, the length of the two fields ($L_B$, $L_E$), the drift after the fields ($D_B$, $D_E$), the charge state of the ion $q$, the electric charge $e$, the kinetic energy $E_{kin}$ and the mass $m$.

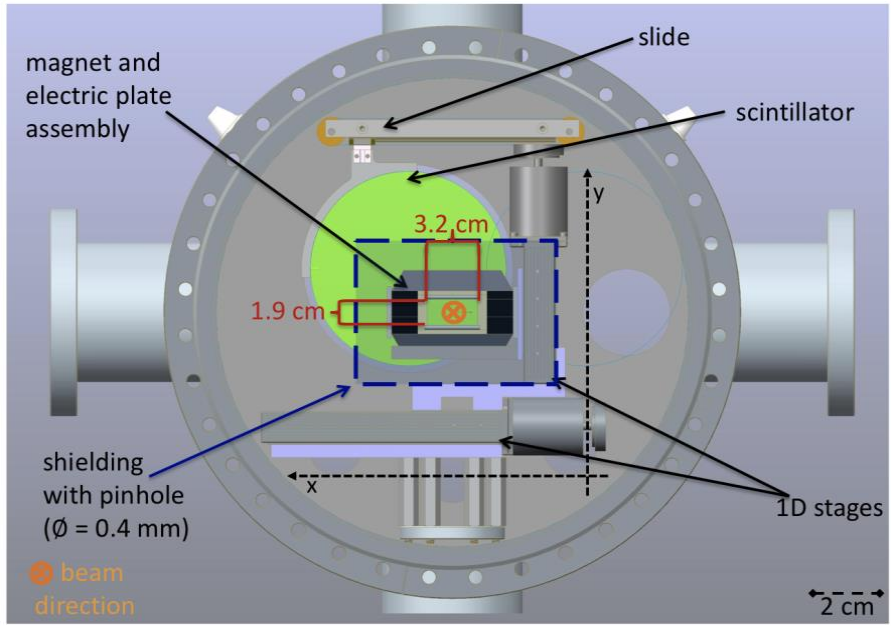

FIG. 2. Two 1D stages allow x and y positioning of the B - E assembly. The scintillator (green) and the B-E assembly can be moved to an "out" position enabling a free line of sight to the target to diagnose the beam and target properties during experiments. The shielding plate with the pinhole ($d_2 = 0.4$ mm) upstream from the assembly is shown as a dashed blue rectangle.

From the modeled field distribution (Opera[10]) the effective magnetic field $B_{eff}$ along the z-axis is calculated following the equation

$$B_{\text{eff}} = \frac{1}{L_B} \int_{-\infty}^{\infty} B_y(z)\, dz = 0.3466\, T \qquad (3)$$

with the geometric length of the magnets $L_B$ and the dominant magnetic field in the y direction $B_y$ that causes the deflection of the particle. The maximal magnetic field $B_{\max}$ is 36 % smaller than $B_{\text{eff}}$ leading to the correlation

$$B_{\text{eff}} \times L_B = 1.36\, B_{\max} \times L_B . \qquad (4)$$

A layer of glass ceramic (Macor) insulates the electric plates from the permanent magnet. A precise HV calibration of the manually controlled potentiometer setting on the front panel of the HV supplies to the measured output voltage, coupled to the simulated electric field in 2D, yields an effective electric field $E_{\text{eff}}$ over the geometric length of the plates $L_E$ as a function of the potentiometer setting:

$$E_{\text{eff}}(pot) = \frac{1}{L_E} \int_{-\infty}^{\infty} E_y(pot, z)\, dz = (3.2672 \times pot - 0.0803)\, \frac{\text{kV}}{\text{cm}} . \qquad (5)$$

For the presented measurements a potentiometer setting of 2.1 was used yielding an electric field of 6.78 kV/cm. The effective electric field is 9 % higher than the maximal simulated field:

$$E_{\text{eff}} \times L_E = 1.09 \cdot E_{\max} \times L_E . \qquad (6)$$

These results emphasize the necessity of modeling of the magnetic and electric field since their distribution and fringing field effects are clearly important to achieve an accurate absolute energy determination.

With the simulated effective magnetic and electric fields it is possible to calculate the expected deflection of the ions, as an example the deflection for 1,000 keV and 500 keV ions is given in TABLE I for both charge states.

TABLE. I. The constant effective magnetic field of 0.3466 T is present over the physical length of the magnet. Similar to that a constant effective electric field of 6.78 kV/cm being present between the electric plates has been calculated. Using these two fields it is possible to calculated the deflection of the particles in x and y, which was done for 1,000 keV and 500 keV and both possible charge states for Helium.

| ion energy [keV] | 1000 | | 500 | |
|---|---|---|---|---|
| charge state [e] | 1+ | 2+ | 1+ | 2+ |
| y deflection (electric) [cm] | 0.91 | 1.82 | 1.82 | 3.64 |
| x deflection (magnetic) [cm] | 1.98 | 3.96 | 2.80 | 5.60 |

## III. EVALUATION ALGORITHM

The original CCD images have a background level of ~90 counts with fluctuations of the order of 5 counts (rms). To efficiently extract the ion positions on the detector an evaluation algorithm has been developed and implemented in Python. Before each set of measurements, a background image is acquired and smoothed using a Gaussian filter (σ = 5 pixels). The filtered background image is then used for background subtraction for every shot in the following measurements. Negative pixel values during background subtractions are set to 0. The region of interest is manually selected by applying a mask to the image. For each pixel the eight adjacent pixels within the mask are tested whether they are above or below the threshold $t$

$$t = \frac{1}{4} \cdot (I_{\min} + \bar{I}) \qquad (7)$$

with the minimal intensity $I_{\min}$ and the average intensity $\bar{I}$ of the pixels within the mask. If at least five surrounding pixels are above the threshold, the pixel is declared to be part of the ion trace, otherwise the pixel value is set to 0. We note that the resulting energy distribution is influenced by the chosen threshold for the trace discrimination. After successful identification of the traces, the Thomson parabola needs to be calibrated by determining the origin of the deflection, corresponding to the detection point of ions for the case $E_{eff} = B_{eff} = 0$. Since the deflection of the ions is a superposition of the deflection caused by the magnetic field and the electric field it is necessary to consider both cases individually for the calibration. For the first calibration measurements, only the deflection of the ions due to the magnetic field is considered. Several measurements without an electric field and with and without the target in the beam path are taken. These measurements result in beam spots lining up on a single line, see green line in Figure 3. For the case of a target in the beam path, separated $He^{1+}$ and $He^{2+}$ traces are observed due to the charge-to-mass ratio sensitivity of the Thomson parabola. The combined traces for the magnetic deflection calibration are fitted, weighted by their intensity, using the linear fit function

$$f_1(x) = -a \cdot x + b_1 \qquad (8)$$

with the fit parameters $a$, slope, and $b$, y-intercept. In a second set of measurements (without the target), the electric deflection calibration is carried out. Here, the electric field is varied for the coexisting fixed magnetic field of the permanent magnets. The beam traces obtained for these measurements also falls on a single line, corresponding to the red line in Figure 4. This line is fitted, weighted by the intensity of the traces, using

$$f_2(x) = -\frac{1}{a} \cdot x + b \qquad (9)$$

with the slope $1/a$ as the value from the previous fit of the magnetic deflection line and the y-intercept $b_1$. The correlation of the slopes of these two fit functions is a result of the orthogonality constraint of these two lines.

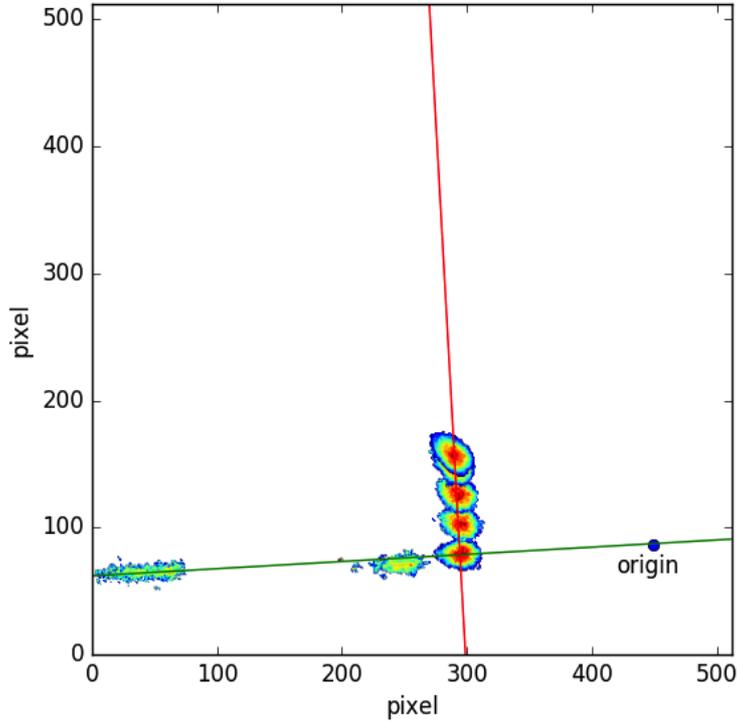

FIG. 3. Initial and transmitted ion traces for no electric field are used to fit the magnetic deflection line (green). The initial beam traces for different electric fields are used to fit the electric deflection line (red). With the average initial energy from the electric deflection, the position of the origin along the magnetic deflection line is calculated.

An average beam energy $\bar{E}$ can be calculated for the data from the second measurement set (variation of the electric field). Using $\bar{E}$ and the intersection point of the two lines, one can calculate the point on the magnetic deflection line that corresponds to no deflection or infinite beam energy. This point cannot be directly measured since the magnetic field created by the permanent magnets is fixed. Once the origin has been determined, it is possible to calculate the energy referring to the displacement of the particles traces from the origin, $y_{el}$ and $x_{mag}$, using the charge state $q$ and the ion mass $m$ based on Equations (1) and (2)

$$E_{kin,el}(y_{el}, q) = \frac{q\, e\, E\, L_E\, (D_E + \frac{L_E}{2})}{2\, y_{el}} \text{ and} \qquad (10)$$

$$E_{kin,mag}(x_{mag}, q, m) = \frac{q^2\, e^2\, B^2\, L_B^2\, (D_B + \frac{L_B}{2})^2}{2\, m\, x_{mag}^2}. \qquad (11)$$

The accuracy of this determined energy is limited by the pixel size of the detector, which has a 512 x 512 pixel array with pixel size 19 x 19 µm. The corresponding energy difference to the 19 µm, or 1 pixel, has been calculated for 1,000 keV and 500 keV ion energy and both possible charge states in TABLE III. The maximum energy difference, 12 keV, due to pixel separation is given for 1,000 keV $He^{1+}$ ions, for the other cases it is well below 10 keV. This means, that the detector size will not influence the energy resolution of the Thomson parabola significantly.

TABLE. II. Due to the pixel size of the detector of 19 x 19 µm the minimal detectable energy difference is given by the pixel separation. For an effective magnetic field of 0.3466 T and an effective electric field of 6.78 kV/cm this separation has been determined in units of keV for 1,000 keV and 500 keV ion energy as well as the two different charge states.

| ion energy [keV] | 1000 | | 500 | |
|---|---|---|---|---|
| charge state [e] | 1+ | 2+ | 1+ | 2+ |
| pixel separation [keV] | ~12 | ~6 | ~5 | ~2 |

## IV. EXPERIMENTAL RESULTS

After the successful implementation of the Thomson parabola, two proof-of-principle experiments are conducted. The initial beam is characterized in terms of average initial beam energy and energy distribution to compare experimental results to simulation results from WARP. Energy-loss measurements are performed for 1 µm thick silicon nitride target foils to compare experimental values and simulated values obtained with SRIM[11].

### A. Initial Beam Characterization

The initial beam energy can be independently determined for the electric and magnetic deflection using Equations (10) and (11). In the discussion below, all shots for one electric field setting are averaged to increase the signal to noise ratio. All initial beam images show only one trace, which is identified as $He^{1+}$ ions with no $He^{2+}$ detected. Weighted by the intensity the average beam energy is $\bar{E} = 1,060$ keV. The shot to shot variation is 30 keV, or ~2 %. Compared to the WARP-simulated average energy, $\bar{E}_{WARP} = 1,045$ keV, the two mean energies $\bar{E}$ and $\bar{E}_{WARP}$ differ by 15 keV, or 1.4 %. Projecting the trace intensities onto the respective axes yields the energy distribution of the incident beam. Figure 4 shows the magnetic (blue) and electric (red) energy distributions for an electric field of 7.76 kV/cm with a chosen bin width of 10 keV as well as the energy distribution predicted by WARP (green). Both distributions exhibit a peaked shape, which agrees well with the predicted distribution except for a missing low energy tail that is predicted by WARP. This result can be due to the averaging of the images for a given setting and background subtraction.

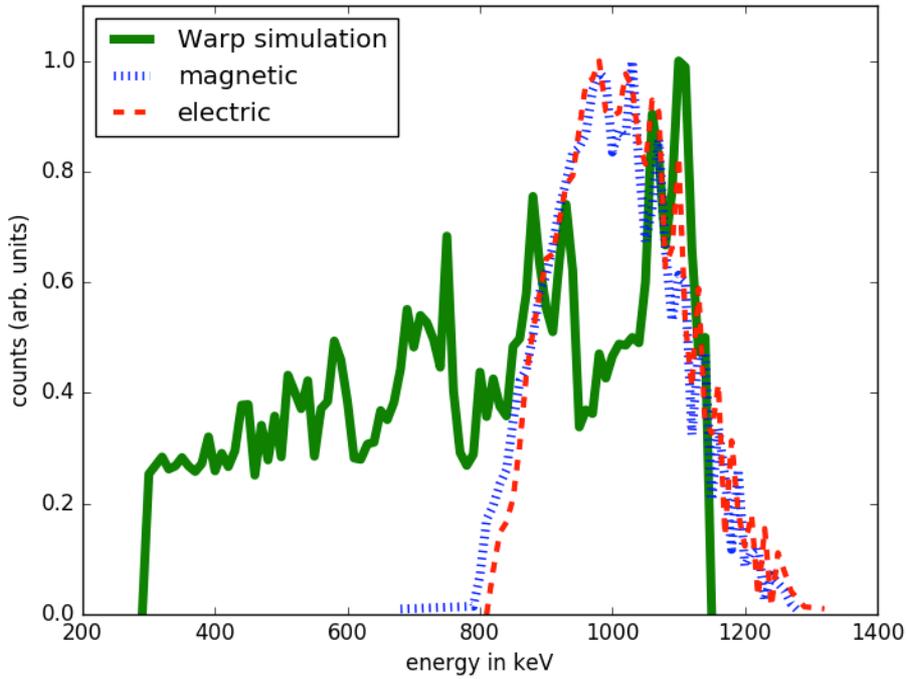

FIG. 4. The energy distribution within the electric (red) and magnetic (blue) deflection at an electric field setting $E_{TP}$ of 7.76 kV/cm show an identical shape. The maxima of both distributions are shifted by 11 keV with a FWHM of 240 keV for the magnetic and electric deflection. Additionally the ion beam energy distribution as predicted by WARP is shown in green with a maximum energy of 1130 keV. It is peaked at a slightly higher energy than the experimental distributions and exhibits a low energy tail which is missing for the experimental distributions.

The 11 keV energy difference between the maxima shows a good consistency for both deflection planes. The full width at half maximum (FWHM), of the data are 240 keV for both deflection planes, almost three times the width of the high energy part of the beam as predicted by WARP simulations (82 keV). The resolution for energies at 1,050 keV is ~140 keV FWHM which is affected by the intrinsic resolution of the Thomson parabola as well as the geometry of the experiment, meaning the two pinholes and the fact that the beam cannot be considered a point source. This assumes a uniform beam that subtends the first aperture with an angular spread constrained by the second aperture ($\approx$ 2.5 mrad). The incident beam profile is peaked and varies by about 30 % across the aperture adding a small correction to the resolution. The energy resolution is directly proportional to the pinhole size.

Using a more sensitive scintillator would improve the signal to noise ratio. A good signal to noise ratio can be beneficial to the trace discrimination method, which could be more successful in selecting only the ion trace rather than including noise in the images. An inefficient background rejection could be the cause of the smooth edge of the energy distribution for lower energies and the rough edge for higher energies. In addition to that the alignment of the Thomson parabola has to be conducted carefully, since a small divergence of the beam due to setup misalignment can result in larger energy shifts, 100 keV energy difference imply a spatial shift of 40 µm.

### B.  Energy-Loss Measurement using Silicon Nitride

The energy-loss of the helium ion beam, 1.6 x 10$^9$ ions per shot and 2 mm FWHM on target, is measured for an amorphous 1um thick silicon nitride (SiN) foil. The experimental result in then compared to the simulated energy-loss obtained with SRIM[11]. For representative simulation results the simulated initial energy distribution from WARP was used as input for the SRIM simulations, assuming a uniform distribution with range (1,045 ± 41) keV. With this simulation input a theoretical energy-loss of 537 keV is obtained. The experimental energy-loss is determined for measurements without an electric field and also for an electric field of 6.78 kV/cm. For each measurement the averaged traces of the transmitted beam are shown in Figure 5. Due to charge exchange and stripping in the target, the transmitted beam shows two charge states, $He^{2+}$ and $He^{1+}$.

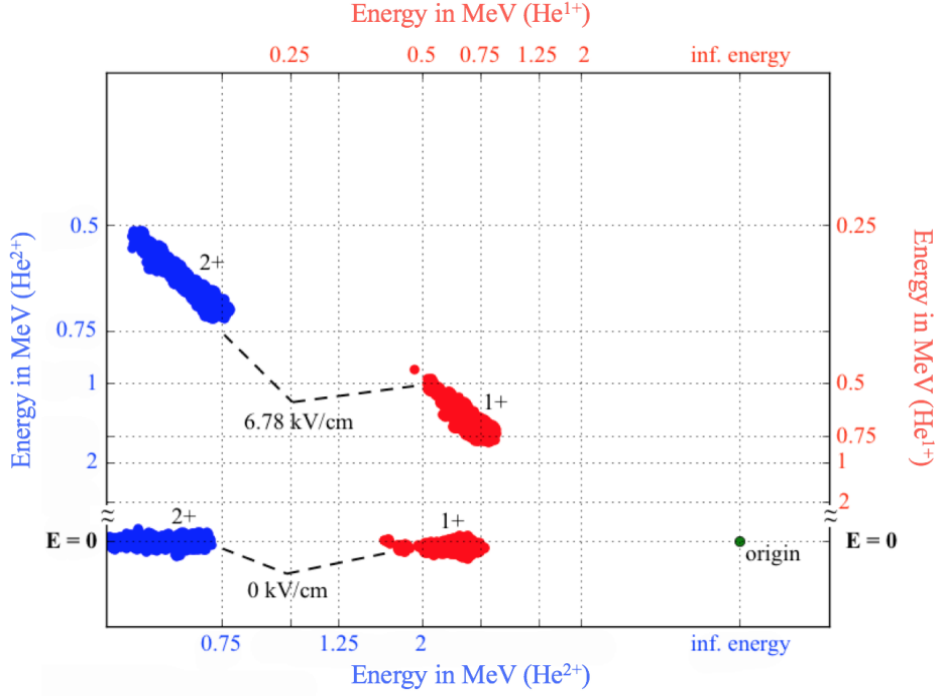

FIG. 5. The averaged traces for the two helium charge states, $He^{1+}$ and $He^{2+}$, are shown for 0 kV/cm (bottom, 10 averaged shots), marked with E = 0 on the y-axis, and 6.78 kV/cm (top, 40 averaged shots) after background subtraction. When there is an electric field present in the Thomson parabola the particles are deflected upwards. For each charge-to-mass ratio the deflected ions are located along parabolas, which intersect at the origin (corresponding to infinite energy).

The deflection for $He^{2+}$ ions is greater than for $He^{1+}$ due to its proportionality to the charge state [see Equations (1) and (2)]. The average transmitted energy for all traces is $\bar{E}_{trans} = (635 \pm 28)$ keV with the error of 28 keV given by the standard deviation of the individual transmitted energies in each beam spot. Due to the low energy resolution of the experiment it was decided to determine the average energy-loss for all settings and both charge states combined. The difference between $\bar{E}_i$ and $\bar{E}_{trans}$ gives an average energy-loss $\bar{E}$ of $(427 \pm 41)$ keV. The fairly high standard deviation value of the transmitted energy and the energy-loss results from energy differences for different field settings (see TABLE III). Here, the energy-loss for the two different electric field settings is shown. For no electric field the energy-loss is only calculated from the magnetic deflection. For the non-zero electric field, the average energy-loss for the electric and the magnet deflection can be determined. The measured energy-loss for no electric field is higher than for an electric field setting of 6.78 kV/cm, and the calculated energy-loss for the electric deflection is higher than the one calculated for the magnetic deflection.

TABLE. III. The averaged transmitted energy values we obtained vary with deflection field settings. For no electric field the average energy-loss value is 10 % higher than values obtained with an applied electric field. Furthermore, for an applied electric field the average energy-loss determined with the electric deflection is 10 % higher than the average energy-loss obtained with magnetic deflection. The average energy-loss value obtained by weighing these three conditions is 427 keV.

| electric field strength [kV/cm] | 0 | 6.78 | 6.78 |
|---|---|---|---|
| deflection type | magnetic | electric | magnetic |
| average energy-loss [keV] | 455 | 431 | 394 |

The SRIM simulated average energy-loss $E_{SRIM}$ of 537 keV is larger than the total measured average energy-loss by 109 keV. Contributing errors include the energy spread of the beam before interacting with the target, the straggling effect of the ions in the target and the resolution of the Thomson parabola. In addition, the energy distribution of the transmitted beam is asymmetric. This asymmetry, coupled with the threshold for the trace selection introduces a bias towards higher transmitted energies. When examining the transmitted traces closely it is obvious that the traces are more intense toward higher energies and are thinner and less intense towards lower energies supporting the thesis of preferred selection of higher energies.

In addition to the energy-loss of the ions the average charge state was also investigated. Weighted by the intensity for each charge state the experimental average charge state is calculated to be (1.6 ± 0.05), which is 6% smaller than the theoretical average charge state of 1.7 using the Schiwietz model[12].

To be able to make more precise measurements and distinguish between the energy-losses for the two charge states it will be necessary to optimize the evaluation algorithm and to improve the intrinsic resolution of the Thomson parabola. This is achievable by exchanging the pinhole with a smaller pinhole. In addition to that the signal to noise ratio can be improved by exchanging the scintillator with a more sensitive scintillator or a multichannel plate detector to facilitate the trace discrimination. Decreasing the gate time of the CCD camera is also a possibility to improve the signal to noise ratio of the measurement, as long as it will be ensured that the gate width is still sufficient to measure the full signal from the scintillator.

## V. CONCLUSION

We have designed and carried out the first experiments with a Thomson parabola for the characterization of the complex beam distribution and for transmission energy-loss experiments on NDCX-II. The measured average kinetic energy of the ion beam distribution agreed well with predictions form WARP particle-in-cell simulations. The measured energy spread at the focal plane was greater than that in the PIC simulations due to resolution effects and limited by the chosen apertures of the Thomson parabola. In a first round of measurements of the energy-loss of helium ions in silicon nitride, the energy-loss was determined to be 428 keV on average, which is 20% less than the SRIM prediction of 537 keV. These results are pointing to instrumental causes of the deviation from background rejection and a systematic correlation between signal strength and ion energy for the low intensity tail of the energy distribution. The initial beam energy was measured within 2 % deviation of the Warp prediction to be 1.06 MeV. Although the average beam energy agrees very well with the simulations, the theoretical beam distribution wasn't reproduced, which can be an effect of the discrimination algorithm as well as detector sensitivity and resolution. Improvements to the resolution of the Thomson parabola will be implemented in future experiments. The aim is to improve the trace selection algorithm as well as the detector resolution for a better signal-to-noise ratio through decreasing the pinhole size and exchanging the detector unit with a more sensitive detector. Additionally, it is aimed to investigate and understand the head versus tail beam dynamics by applying shorter gates to the camera. In the future, the energy-loss in materials at different ion fluxes will be investigated to explore possible fluence effects in a higher ion fluence regime. This work shows that NDCX-II can be used to study basic ion solid interactions.

**ACKNOWLEDGMENTS**

The authors would like to thank Takeshi Katayanagi and Bill Ghiorso for the technical support. This work is supported by the US Department of Energy, Office of Science, under contract DE-AC0205CH11231 (LBNL), DE-AC52- 07NA27344 (LLNL), and DE-AC02-09CH11466 (PPPL).